\documentclass[aps,prl,floatfix,twocolumn,showpacs,10pt]{revtex4-2}

\usepackage{graphicx}
\usepackage{dcolumn}
\usepackage{bm}
\usepackage{amssymb}
\usepackage{amsmath} 
\usepackage{rotating} 
\usepackage{xcolor}
\usepackage[hidelinks]{hyperref}

\begin{document}

\title{Representability for Quantum Theory beyond Particle-Number Conservation}

\author{David A. Mazziotti}

\email{damazz@uchicago.edu}
\affiliation{Department of Chemistry and The James Franck Institute, The University of Chicago, Chicago, IL 60637}%

\date{Submitted December 10, 2025\textcolor{black}{; Revised February 2, 2026}}


\begin{abstract}
Representability determines when a two-particle reduced density matrix (2-RDM) corresponds to a physical quantum state, enabling many-particle quantum calculations with 2-RDMs rather than the wave function. In this Letter, we present a solution of the representability problem for quantum systems without particle-number conservation. The physically allowed set of 2-RDMs can be characterized from a geometrically `orthogonal' set, the polar cone. We derive explicit linear equations for the two-body operators in the polar cone---the intersection of the $p$-positive cone with the two-body operator space---to obtain a systematic hierarchy of representability conditions that do not depend on higher RDMs or the wave function. Moreover, by augmenting these conditions with the particle-number variance, we obtain a unified framework for treating both particle-number-conserving and nonconserving systems. We illustrate with a spin system and molecular H$_4$.
\end{abstract}


\maketitle

{\em Introduction---}Although the wave function is central to most approaches to the many-body problem, it often contains far more information than required and, without approximation, typically scales exponentially with system size~\cite{Mazziotti2007, Coleman2000}. For many-particle systems with two-body interactions, however, the energy and all one- and two-particle properties can be expressed as an affine functional of the two-particle reduced density matrix (2-RDM)~\cite{Coleman1963}. The ground-state energy cannot be directly computed as a functional of the 2-RDM without the wave function unless the 2-RDM satisfies nontrivial representability conditions for its correspondence to a physical quantum state~\cite{Coleman1963, Garrod1964, Kummer1967, Erdahl1978, Erdahl1989, Coleman2000, Mazziotti2007, Mazziotti2012b, Mazziotti.2023}. Such calculations have been realized through approximate representability conditions with significant applications to strong correlated molecules and materials~\cite{Erdahl2000, Nakata2001, Mazziotti2001c, Mazziotti2002b, Zhao2004, Mazziotti2004a, Mazziotti2005, Cances2006, Mazziotti2006b, Erdahl2007, Braams2007, Gidofalvi2008, Shenvi2010, Mazziotti2011, Verstichel2012, Baumgratz2012, Poelmans2015, Mazziotti2016, Alcoba2018, Rubio-Garcia2019, Head-Marsden2020, Haim2020, Han2020, Mazziotti2020, Li2021, Knight2022, Kull.2024, Gao.2025dgs, Knight.2025, Schouten.2025} including recent applications from conduction in glassy polymers~\cite{Xie2022} to exciton condensation in molecular materials~\cite{Safaei2018, Schouten.2023, Schouten.2024, Torres.2024} (see also~\cite{Schilling2021, Benavides-Riveros2020, Schilling2019, Piris2021, Piris2017a, Gibney2022a, Liebert.2025axw} for connections with one-particle RDM functional theories). While a systematic hierarchy of such constraints has been established for fixed particle number---the $N$-representability conditions~\cite{Mazziotti2012b, Mazziotti.2023}---a comparably general framework has not been developed for the number-nonconserving case.

In this Letter, we present the representability conditions of the 2-RDM beyond particle-number conservation\textcolor{black}{, i.e., the conditions for representability by a (possibly mixed) density operator on Fock space.}  The set of representable 2-RDMs can be defined directly---without wave functions---by a geometrically `orthogonal' set known as the polar cone~\cite{Coleman2000, Kummer1967, Erdahl1978, Mazziotti2012b}. We characterize the two-body operators in the polar cone, which we show to be the intersection of the positive cone in Fock space with the two-body operator space. Specializing to fermions (though the approach extends naturally to bosons and spin systems), we decompose arbitrary operators into irreducible $p$-body components in a Majorana-like basis and enforce the cancellation of all contributions beyond two-body. This procedure yields a systematic hierarchy of representability conditions---the $(2,p)$-positivity conditions---without assuming particle-number conservation. We further show that this same hierarchy applies to the fixed-$N$ case when supplemented by the variance of the number operator, thereby unifying the number-conserving and number-nonconserving theories within a single constructive framework. Illustrative applications are presented for a pairing Hamiltonian and molecular H$_{4}$.

{\em Theory---} For quantum many-particle systems with pairwise interactions, we can write the ground-state energy directly in terms of two-particle operators~\cite{Mazziotti2007, Coleman2000},
\begin{equation}
E = \mathrm{Tr}\!\left(\hat H\, {}^{2}\!\hat D\right),
\end{equation}
where $\hat H$ and ${}^{2}\!\hat D$ are the Hamiltonian and the two-particle reduced density operators \textcolor{black}{(2-RDO)}.  \textcolor{black}{Let $\{\hat\Gamma_\alpha\}$ denote a spanning set for the two-body operator space $\mathcal H_2$ on Fock space.}  The matrix elements of the 2-RDM are
\begin{equation}
{}^{2}D^{ij}_{kl} = \mathrm{Tr}\!\left(\hat\Gamma^{ij}_{kl}\, {}^{2}\!\hat D\right).
\label{eq:rdm2}
\end{equation}
\textcolor{black}{In the particle-number–conserving case, a standard choice for $\hat\Gamma^{ij}_{kl}$ is ${\hat a}^{\dagger}_{i} {\hat a}^{\dagger}_{j} {\hat a}_{l} {\hat a}_{k}$, which agrees with the conventional definition of the 2-RDM.}  Variational calculation of ${}^{2}\!\hat D$, however, requires enforcing constraints to ensure that it is representable by a physical quantum state---the representability conditions.  While these constraints are well characterized in the fixed-$N$ case through the $N$-representability framework~\cite{Mazziotti2012b, Mazziotti.2023}, a similar formulation has not been developed in the number-nonconserving case.

The representability conditions for 2-RDOs can be formally expressed by the bipolar theorem~\cite{Coleman2000, Kummer1967}
\begin{equation}
\mathrm{Tr}\!\left({}^{2}\!\hat B\, {}^{2}\!\hat D\right)\ge 0
\qquad\text{for all } {}^{2}\!\hat B\in{}^{2}P^{*},\; {}^{2}\!\hat D\in {}^{2}P,
\end{equation}
where ${}^{2}P$ is the cone of representable 2-RDOs and ${}^{2}P^{*}$ is its polar cone.  Consequently, the representability problem reduces to identifying the two-body operators ${}^{2}\!\hat B$ in ${}^{2}P^{*}$.  Because every representable ${}^{2}\!\hat D$ arises from a positive operator $\hat D\in P$ where $P$ is the cone of all positive semidefinite operators on Fock space, the inequality above is equivalent to
\begin{equation}
\mathrm{Tr}\!\left({}^{2}\!\hat B\, \hat D\right)\ge0
\qquad\text{for all } {}^{2}\!\hat B\in{}^{2}P^{*},\; \hat D\in P,
\end{equation}
implying that ${}^{2}\!\hat B$ itself must lie in the positive cone $P$.  By definition ${}^{2}\!\hat B$ is also a two-body operator, and hence, the polar cone of representability conditions on the 2-RDO is the intersection
\begin{equation}
{}^{2}P^{*} = P \cap \mathcal{H}_{2},
\end{equation}
the set of all operators that are both two-body and positive semidefinite on Fock space~\cite{Erdahl1978}.

To make this intersection constructive, we introduce a hierarchy of conditions.  We define the $p$-positive cone $P_{p}$ as the convex cone generated by Hermitian squares of $p$-body Majorana polynomials,
\begin{equation}
P_{p} = \mathrm{cone}\Big\{\,
\hat O_i \hat O_i^{\dagger}
\;:\;
\hat O_i = \hat\gamma_{\mu_1}\hat\gamma_{\mu_2}\cdots\hat\gamma_{\mu_p}
\,\Big\},
\end{equation}
where each $\hat O_i$ is a product of $p$ Majorana operators $\hat\gamma_{\mu}$.  Restricting the intersection to $P_{p}$ defines the approximate polar cone
\begin{equation}
{}^{2}P^{*}_{p} = P_{p} \cap \mathcal{H}_{2},
\end{equation}
which yields the $(2,p)$-positivity conditions.

To make the conditions explicit, we employ the canonical decomposition of operators on Fock space into irreducible body-order components~\cite{Erdahl1978, Bach.2019},
\begin{equation}
\hat B = \sum_{p=0}^{r} \hat B^{(p)},
\end{equation}
where $\hat B^{(p)}$ denotes the $p$-body part in a one-fermion basis of rank $r$.  This decomposition generalizes the unitary decomposition~\cite{Coleman1980, AuChin1983} used in the fixed-$N$ derivation of representability~\cite{Mazziotti.2023}.  Expressing a general operator of body order $\le p$ as a polynomial in Majorana-like operators $\hat \gamma_{\mu}$---linear combinations of creation and annihilation operators into Hermitian and anti-Hermitian operators---allows us to identify all $q$-body operators for $q>2$, as shown by Erdahl~\cite{Erdahl1978}.  We enforce membership in $\mathcal{H}_{2}$ by setting the coefficients of all higher-than-two-body terms of an operator in $P_{p}$ to zero, which produces a linear system defining the admissible two-body operators.  The resulting two-body operators ${}^{2}\!\hat B \in {}^{2}P^{*}_{p}$ form a systematic number-nonconserving hierarchy of representability conditions---$(2,p)$-positivity conditions---that converges to the exact polar cone as $p$ increases.

Both number-conserving and number-nonconserving fermionic states are treatable within a single representability framework.  If we constrain the particle-number variance to vanish
\begin{equation}
\mathrm{Tr}\!\left[(\hat N - N)^{2}\,{}^{2}\!\hat D\right] = 0,
\end{equation}
we recover the fixed-$N$ sector.


\begin{figure}[ht!]

\includegraphics[scale=0.4]{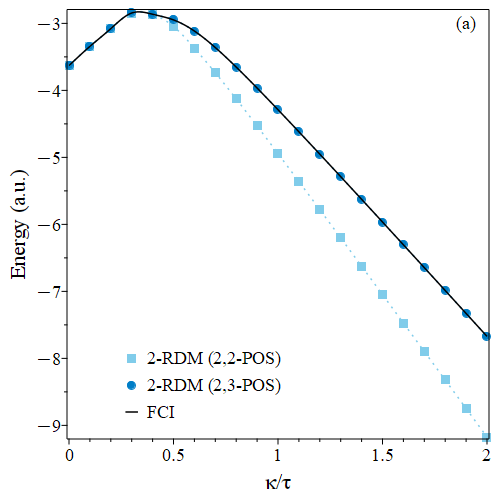}

\includegraphics[scale=0.4]{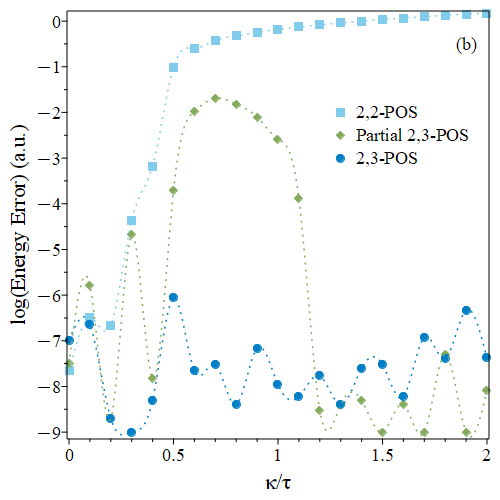}

\caption{(a) Ground-state energy and (b) base-10 energy error relative to FCI of the six-orbital pairing ring Hamiltonian from V2RDM with
(2,2)-positivity, partial (2,3)-positivity, and full (2,3)-positivity.}

\label{f:pair}

\end{figure}

While the representability constraints can be added to the energy minimization, we solve an equivalent dual semidefinite program (SDP):
\begin{align}
\max_{E, \, \lambda \ge 0,\,{}^{p}\!\hat X \succeq 0}\; E \qquad & \\
\text{such that~~}
\hat H - E - {}^{2}\!\hat B - \lambda(\hat N - N)^{2} &= 0, \\
{}^{p}\!\hat X - {}^{2}\!\hat B &= 0.
\end{align}
We maximize the energy such that $\hat H - E - \lambda(\hat N - N)^{2}$ is expressible as an operator ${}^{2}\!\hat B \in {}^{2}P^{*}_{p}$, which enforces the $(2,p)$-positivity conditions and, if $\lambda > 0$, the vanishing of the particle-number variance.  We fix $\lambda=0$ to recover the fully number-nonconserving formulation.  As in the number-conserving theory~\cite{Mazziotti.2023}, the Lagrange multiplier associated with the first constraint is the 2-RDO expressed in the Majorana-like operator basis with the 2-RDM computable by Eq.~(\ref{eq:rdm2}).

{\em Results---}We apply the variational 2-RDM (V2RDM) method with the number-nonconserving $(2,p)$-positivity conditions to two representative systems: a six-orbital fermionic ring that exhibits particle-number nonconservation and the symmetric dissociation of linear H$_4$, which remains strictly number conserving but develops strong multireference correlation as the bonds stretch.  The dual semidefinite programs are solved with the boundary-point algorithm implemented in Ref.~\cite{Mazziotti2011}.  In this algorithm, without exploiting any symmetry or sparsity, the V2RDM method with the $(2,2)$- and $(2,3)$-positivity conditions scales as $r^{6}$ and $r^{9}$ floating-point operations and requires $r^{4}$ and $r^{6}$ storage, respectively. \textcolor{black}{In the number-nonconserving case of the fermionic ring, full configuration interaction (FCI) corresponds to exact diagonalization of the Hamiltonian in the full Fock space, including all particle-number sectors, rather than restriction to a fixed-$N$ subspace.}  For H$_4$, treated in the minimal Slater-type-orbital (STO-3G) basis set~\cite{Hehre1969}, electron integrals as well as energies from second-order many-body perturbation theory (MP2), coupled cluster with single, double, and perturbative triple excitations [CCSD(T)], and FCI are obtained with the Quantum Chemistry Package in Maple~\cite{QCT2022}.  \textcolor{black}{Minimal system sizes and basis sets are chosen to enable direct comparison with exact FCI energies; we do not yet exploit spin and spatial symmetries, sparsity, or low-rank structure in the V2RDM calculations, nor in the number-nonconserving (fermionic-ring) FCI reference.}

The six-orbital ring Hamiltonian is
\begin{align}
\hat H &=
\frac{\mu}{2}
\left( r \hat I + \sum_{i=1}^{r} \hat p_i \hat m_i \right)
\label{eq:H1}
\\
&
+ \sum_{i=1}^{r}
\left[
\frac{\tau}{2}
  (\hat p_i \hat m_{i+1} + \hat p_{i+1} \hat m_i )
+
\frac{\Delta}{2}
  (\hat p_i \hat m_{i+1} - \hat p_{i+1} \hat m_i )
\right]
\nonumber
\\
&
+ \kappa \sum_{i=1}^{r}
\left[
(\hat p_i \hat p_{i+1})(\hat m_{i+2}\hat m_{i+3})
+
(\hat p_i \hat m_{i+1})(\hat p_{i+2}\hat m_{i+3})
\right].
\nonumber
\end{align}
with all indices taken modulo $r$, and where $\hat p_i = \hat a^{}_i + \hat a_i^\dagger$ and $\hat m_i = \hat a^{}_i - \hat a_i^\dagger$ are Majorana-like fermionic operators.  This model generalizes the Nambu Bardeen-Cooper-Schieffer (BCS) Hamiltonian~\cite{Bardeen.1957,Nambu.1960} by including more general two- and four-body interactions while still explicitly breaking particle-number conservation.  The model parameters are chosen as $\mu=-0.2$, $\tau=0.5$, $\Delta=0.5$, and $\kappa/\tau$ varied between $0$ and $2$.  The parity of the ground state changes from odd to even near $\kappa/\tau \approx 0.5$.

The ground-state energy as a function of $\kappa/\tau$ from V2RDM with $(2,2)$-positivity and $(2,3)$-positivity together with FCI is shown in Fig.~1(a).  Both levels of positivity agree closely with FCI for $0 \le \kappa/\tau \le 0.5$.  Beyond $\kappa/\tau = 0.5$, where the ground-state parity changes, the $(2,2)$-positivity energies deviate significantly from FCI, while the $(2,3)$-positivity curve continues to track FCI across the entire interaction range.  The base-10 logarithm of the energy error relative to FCI is shown in Fig.~1(b) for $(2,2)$-positivity, partial $(2,3)$-positivity, and full $(2,3)$-positivity. While $(2,2)$-positivity exhibits errors of order $10^{-1}$~a.u.\ for $\kappa/\tau \gtrsim 0.5$, full $(2,3)$-positivity maintains errors below $10^{-6}$~a.u.\ for all $\kappa/\tau$.  The partial $(2,3)$-positivity conditions---the number-nonconserving analogs of the T1 and T2 conditions~\cite{Zhao2004, Mazziotti2005, Mazziotti2016, Mazziotti2020, Erdahl1978} obtained by retaining only the even particle-hole symmetric three-body submatrices that reduce to two-body operators---show intermediate accuracy with errors of order $10^{-3}$–$10^{-2}$ for $\kappa/\tau \approx 0.5$–$1.2$.  The enhanced accuracy of the full $(2,3)$-positivity conditions demonstrates that computing the full intersection of the three-positive cone with the two-particle space introduces important additional $N$-representability conditions on the 2-RDM beyond the T1 and T2 analogs.


\begin{figure}[htp!]

\includegraphics[scale=0.4]{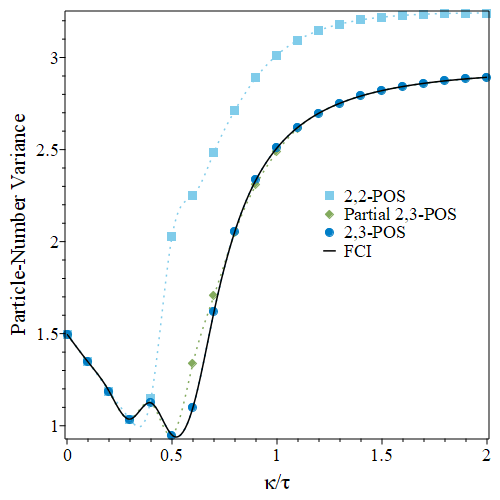}

\caption{Particle-number variance of the ground state of the pairing ring Hamiltonian from V2RDM with (2,2)-positivity, partial (2,3)-positivity, and full (2,3)-positivity, compared with FCI.}

\label{f:pairv}

\end{figure}

Figure~2 shows the particle-number variance as a function of $\kappa/\tau$ from V2RDM with $(2,2)$-positivity, partial $(2,3)$-positivity, and full $(2,3)$-positivity, along with FCI.  The nonzero variance demonstrates that the ground state of the ring Hamiltonian is number non-conserving.  The variance from $(2,2)$-positivity becomes too large for $\kappa/\tau \gtrsim 0.5$, while the variance from full $(2,3)$-positivity is indistinguishable from FCI for all $\kappa/\tau$.  As in the energy comparison, the partial $(2,3)$-positivity results follow FCI near equilibrium but depart from the exact results for $\kappa/\tau \approx 0.5$–$1.2$.


\begin{figure}[htp!]

\includegraphics[scale=0.4]{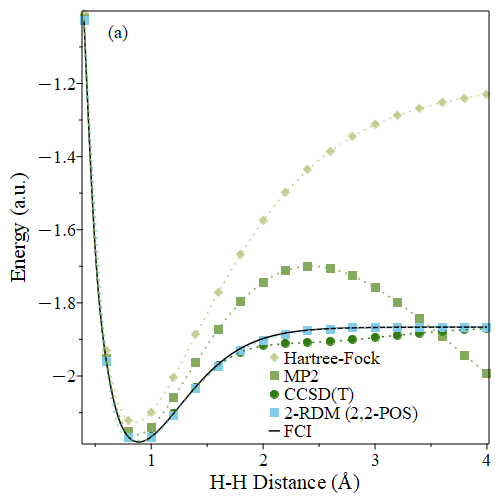}

\includegraphics[scale=0.4]{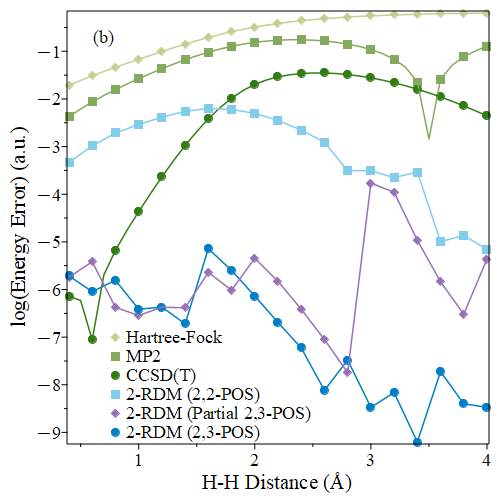}

\caption{(a) Potential energy curve and (b) base-10 energy error relative to FCI for the symmetric dissociation of linear H$_{4}$ from Hartree–Fock, MP2, CCSD(T), and V2RDM with (2,2)- as well as partial and full (2,3)-positivity.}

\label{f:h4}

\end{figure}

We next consider the dissociation of linear H$_4$ with all adjacent H–H bonds stretched equally.  Figure~3(a) shows the potential energy curve from Hartree–Fock, MP2, \textcolor{black}{CCSD(T)}, and V2RDM with $(2,2)$-positivity together with FCI. In this case, the number-operator constraint is employed to select the number-conserving solution within the number-nonconserving variational framework.  The $(2,2)$-positivity curve agrees well with FCI even in the dissociation region that exhibits strong multireference correlation.

The absolute energy errors relative to FCI on a base-10 logarithmic scale from Hartree–Fock, MP2, CCSD(T), and V2RDM with $(2,2)$-positivity, \textcolor{black}{partial $(2,3)$-positivity}, and $(2,3)$-positivity are presented in Fig.~3(b). The CCSD(T) errors range from $10^{-7}$ to $10^{-2}$~a.u.\ and increase significantly as the bonds stretch and the molecule becomes more correlated.  The $(2,2)$-positivity errors range from $10^{-5}$ to $10^{-2}$~a.u., while the $(2,3)$-positivity conditions yield errors between $10^{-9}$ and $10^{-5}$~a.u.  Partial $(2,3)$-positivity matches full $(2,3)$-positivity near equilibrium but approaches the lower accuracy of $(2,2)$-positivity in the dissociation limit, again indicating that the full intersection provides essential representability conditions not captured by the T1 and T2 analogs.

{\em Discussion and Conclusions---}We have presented a solution of the representability problem for quantum theory beyond number conservation.  The operators whose trace against the 2-RDO define the representability conditions are characterized by the intersection of the positive cone in Fock space with the two-body operator space. We derive a system of linear equations for these operators by employing the canonical decomposition of arbitrary fermionic operators into irreducible $p$-body components, which generalizes the unitary decompositions used previously in the solution of the $N$-representability problem~\cite{Mazziotti.2023}. By supplementing the derived representability conditions with the variance of the number operator, we also provide a complementary approach to solving the $N$-representability problem in the fixed-$N$ sector, yielding a unification of the number-conserving and number-nonconserving theories within a single computational framework. The resulting representation enables the formulation of the 2-RDM without higher RDMs as a practical semidefinite program. In general, because the solution of the representability problem is known to be Quantum--Merlin--Arthur (QMA) complete~\cite{Liu2007}, the worst-case scaling of the hierarchy reflects the exponential scaling of the full Fock-space metric, but in practice the constraints can exploit the entanglement complexity of a given system, which often leads to rapid convergence with the level $p$ of the $(2,p)$-positivity conditions.  \textcolor{black}{In practice, the computational cost can be substantially reduced by exploiting spin and spatial symmetries, sparsity, and low-rank structure in the constraint maps and semidefinite variables, as has been demonstrated in number-conserving V2RDM calculations.  For example, exploiting low-rank structure in the T2 condition has been shown to reduce the dominant computational cost from $O(r^{9})$ to $O(r^{6})$, enabling applications to substantially larger systems~\cite{Mazziotti2016, Mazziotti2020}.}

The generalized 2-RDM theory, applicable to both number-conserving and number-nonconserving states and implemented through a common SDP framework, has been applied to a pairing ring Hamiltonian and the symmetric dissociation of linear H$_4$. \textcolor{black}{The pairing ring provides a nontrivial test for particle-number–nonconserving representability, while linear H$_4$ provides a well-known molecular benchmark with available FCI results that exhibits strong static correlation.}  In both cases, the $(2,3)$-positivity conditions improve upon the number-nonconserving analogs of the partial $(2,3)$-conditions (the T1 and T2 conditions), demonstrating that the complete hierarchy of number-nonconserving $(2,p)$-positivity conditions rapidly generates representability constraints beyond the well-known D, Q, G, T1, and T2 conditions. Because the intersection of the $p$-positive cone with the two-body space is in general nonchordal~\cite{Grone.1984, Fawzi.2015}, the resulting system of equations cancels the higher-than-two-body components in a nontrivial way, producing both rank-1 and non-rank-1 representability conditions. The solution does not directly depend on higher RDMs, thereby providing a constructive solution to the representability problem for the two-particle sector. \textcolor{black}{The present work demonstrates, as illustrated here on representative benchmarks, that this approach provides an alternative and highly accurate many-electron framework for both number-conserving and number-nonconserving systems, with broad potential for advancing the description of strongly correlated quantum systems.}

\begin{acknowledgments}

D.A.M. gratefully acknowledges the U.S. National Science Foundation Grant No. CHE-2155082.

\end{acknowledgments}

\noindent {\em Data availability---} The data that support the findings of this article are openly available in Ref.~\cite{Data}.

\bibliography{SFDM-25}

\end{document}